\documentclass[
reprint,
superscriptaddress,
showpacs,
nofootinbib,
amsmath,amssymb,
aps,
prl,
floatfix,
longbibliography,
footinbib
]{revtex4-2}

\usepackage{graphicx}
\usepackage{dcolumn}
\usepackage{float}
\usepackage{booktabs}
\usepackage{bm}
\usepackage{bbm}
\usepackage{mathrsfs}
\usepackage{mathtools}
\usepackage{xcolor}
\usepackage{hyperref}
\hypersetup{colorlinks=true,citecolor=blue,urlcolor=magenta,linkcolor=magenta}

\newcommand{\safeincludegraphics}[2][]{%
	\IfFileExists{#2}{\includegraphics[#1]{#2}}{%
		\fbox{%
			\parbox[c][0.24\textheight][c]{0.46\textwidth}{%
				\centering
				\vspace{2mm}
				Placeholder for\\[1mm]
				\texttt{#2}
				\vspace{2mm}
			}%
		}%
	}%
}

\begin{document}
	
\title{Kibble--Zurek Scaling in the Dicke Model at Mesoscopic Scales}
	
\author{Haowei Li}
\email{hwliphys@gmail.com}
\affiliation{Institute for Advanced Study, Tsinghua University, Beijing 100084, China}
\affiliation{Beijing Key Laboratory of Cold Atom Quantum Computation, Tsinghua University, Beijing 100084, China}

\author{Hanteng Wang}
\email{hantengwang.physics@gmail.com}
\affiliation{Institute for Advanced Study, Tsinghua University, Beijing 100084, China}
\affiliation{Beijing Key Laboratory of Cold Atom Quantum Computation, Tsinghua University, Beijing 100084, China}
	
\begin{abstract}
The Dicke model is a paradigmatic setting for collective light–matter physics and the superradiant phase transition. Yet extracting the critical exponents is challenging at experimentally accessible mesoscopic sizes, due to the slow divergence of the correlation time under all-to-all coupling and a photon-loss-driven crossover to a distinct dissipative universality class. Here, we perform a large-$N$ analysis that identifies distinct coherent and dissipative fixed points for the closed and open Dicke models. We then develop a unified mesoscopic scaling framework that incorporates the leading irrelevant correction and, going beyond static and spectral probes, brings ramping dynamics under the same scaling description. It recovers the corresponding exponents, verifies Kibble–Zurek scaling, and clarifies how finite size, dissipation, and speed compete in the ramping dynamics. Our work thus establishes a unified framework for resolving static and dynamical critical scaling in closed and open quantum systems, with broader applicability to mesoscopic systems with long-range interactions.
\end{abstract}
\maketitle

{\it Introduction.---}
Collective light--matter systems realize long-range interactions among quantum emitters mediated by optical modes~\cite{Ritsch2013,Landig2016,Vaidya2018}. The Dicke model captures this physics as a paradigmatic model of superradiance in both coherent and dissipative cavity-QED settings~\cite{Dicke1954,Kirton2019}. The model exhibits a $\mathbb{Z}_2$ symmetry-breaking transition~\cite{HeppLieb1973,WangHioe1973,EmaryBrandes2003PRL}, near which the low-energy and long-time behavior is governed by universal critical exponents that dictate how observables scale with tuning parameters~\cite{Cardy1996,Sachdev2011}. Extracting these exponents experimentally is a central way to identify the underlying universality class, as has been done in many condensed-matter systems and, more recently, in quantum simulators with a finite number of particles~\cite{Keesling2019,Ebadi2021,King2022DWaveIsingChain,King2023DWaveSpinGlass,Li2023,Zhang2025NearCriticalKZ}.

Despite the simplicity of the Dicke Hamiltonian, a quantitative extraction of its critical exponents has remained challenging. Early cavity-QED realizations reached large atom numbers, but spatial inhomogeneity of the cavity mode drove the light--matter coupling far from the ideal collective limit~\cite{Baumann2010,Brennecke2013,Klinder2015}. Recent tweezer arrays offer nearly uniform couplings, but only at mesoscopic atom numbers~\cite{Seubert2025SubwavelengthCavityArray,Ho2025MesoscopicArray,Wang2025CavityArray,DeSantis2026CavityRydbergArray}. In this regime, all-to-all coupling makes the critical correlation time grow slowly with $N$~\cite{LiebRobinson1972,HastingsKoma2006,Maghrebi2016Causality,Defenu2023}, so finite-size corrections readily obscure its asymptotic scaling and bias estimates of the dynamical exponent~\cite{Li2023,Ho2025MesoscopicArray,De2025}. Unavoidable photon loss~\cite{Dimer2007,Baumann2010} is a relevant perturbation that drives the system toward a distinct dissipative universality class~\cite{Nagy2011,Konya2012,Torre2013}; at mesoscopic sizes and finite loss, the crossover between the coherent and dissipative limits can be broad, so the apparent exponents need not coincide with those of either asymptotic fixed point.

\begin{figure}[t]
    \centering
    \safeincludegraphics[width=0.48\textwidth]{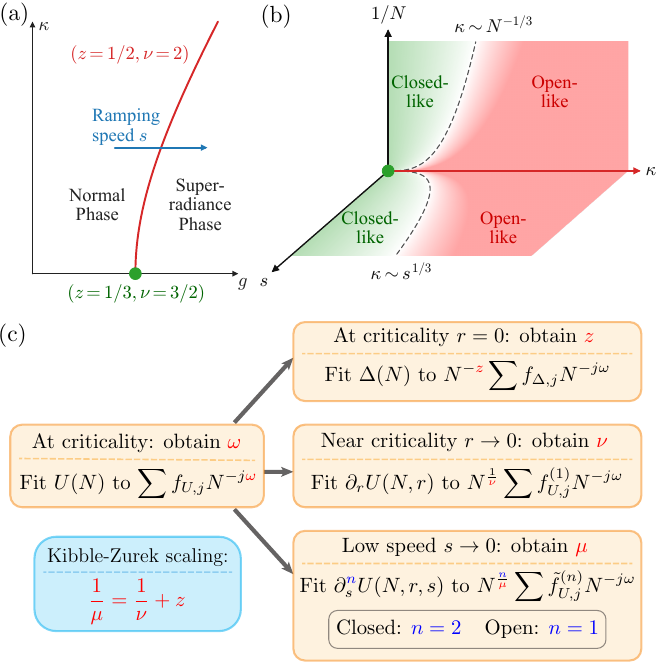}
    \caption{(a) Phase diagram of the Dicke model in the plane of light-matter coupling $g$ and dissipation $\kappa$. Two distinct universality classes are identified at large $N$: closed (green) and open (red). The ramp protocol is defined by the ramping speed $s$. (b) Crossover regions set by size $N$, dissipation $\kappa$, and speed $s$ in the ramping dynamics. (c) Scaling protocol at mesoscopic scales that encodes the irrelevant exponent $\omega$. The exponents $\nu$, $z$, and $\mu$ are then extracted independently, and the KZ relation $1/\mu=1/\nu+z$ is tested.}
    \label{fig:fig1}
\end{figure}

Critical properties can be probed beyond the static and steady-state cases. Under a finite-speed ramp through the transition, Kibble--Zurek (KZ) scaling links the critical exponents to the universal dynamical response~\cite{Kibble1976,Zurek1985,ZurekDornerZoller2005,Polkovnikov2005,Dziarmaga2010,RossiniVicari2021}, a relation verified across diverse systems~\cite{Ulm2013,Navon2015,Keesling2019,Zhang2025NearCriticalKZ,King2023DWaveSpinGlass}. For the Dicke model, the absence of spatial structure eliminates the usual domain-wall picture. Nevertheless, dynamical scaling laws have
been formulated in the asymptotic large-$N$ regime for various closed all-to-all models~\cite{Acevedo2014DynamicalScaling,Defenu2018DynamicalCriticalScaling,Xue2018DrivenCriticalDynamics}. At mesoscopic sizes, however, strong finite-size drift obscures the asymptotic scaling, and a controlled extraction of the exponents from experimentally accessible systems remains lacking. Moreover, the signatures of the coherent and dissipative fixed points become intertwined with the finite ramp speed, making the dynamical critical response difficult to characterize~\cite{Puebla2020,Jara2024} and calling for a protocol that can disentangle these effects.

In this Letter, we develop a unified mesoscopic scaling theory for Dicke superradiant criticality that treats static and driven scaling, as well as coherent and dissipative criticality, on the equal footing. Using a large-$N$ field theory, we identify distinct universality classes for the coherent and dissipative fixed points [Fig.~\ref{fig:fig1}(a)]. 
Because finite size, photon loss, and finite ramp speed introduce competing dynamical time scales, broad crossover regimes emerge in which neither fixed-point response is cleanly visible [Fig.~\ref{fig:fig1}(b)]. To overcome this difficulty, we construct a finite-size scaling protocol that retains the leading irrelevant correction~\cite{Wegner1972,Campostrini2014} and extracts the thermodynamic exponents from mesoscopic static, spectral, and ramped data [Fig.~\ref{fig:fig1}(c)]. This analysis recovers the large-$N$ exponents and verifies the KZ relation in both cases, establishing universal driven scaling in a fully connected model without spatial domain formation.

{\it Dicke model: closed and open.---}
We consider the Dicke model, in which $N$ two-level atoms with energy splitting $\omega_z$ couple to a single cavity mode $\hat a$ of frequency $\omega_0$. The Hamiltonian reads~\cite{Dicke1954}
\begin{equation}
	\hat H=\omega_0 \hat a^\dagger \hat a+\omega_z \hat J_z+\frac{g}{\sqrt N}(\hat a+\hat a^\dagger)\hat J_x,
	\label{eq:Dicke_H}
\end{equation}
where $\hat J_\alpha=\frac12\sum_{i=1}^N\sigma_i^\alpha$ for $\alpha=x,z$, and $g$ is the light-matter coupling strength. The cavity may additionally undergo photon loss at a rate $\kappa$. For $\kappa=0$, the model is closed and evolves coherently; for $\kappa\neq0$, it is open and evolves dissipatively according to the Lindblad equation~\cite{Dimer2007}
\begin{equation}
	\dot{\hat\rho} = -\mathrm{i}[\hat H,\hat\rho] + \kappa \left(2\hat a\hat\rho \hat a^\dagger-\hat a^\dagger \hat a\hat\rho-\hat\rho \hat a^\dagger \hat a\right).
	\label{eq:lindblad}
\end{equation}
In the thermodynamic limit, the model undergoes a superradiant phase transition. The $\mathbb{Z}_2$ symmetry-breaking phase is characterized by a nonzero order parameter, $\mathrm{Tr}(\hat\rho\,\hat x)\neq0$, where $\hat x=(\hat a+\hat a^\dagger)/\sqrt{2\omega_0}$, and the critical point is located at $g_c=\sqrt{(\kappa^2+\omega_0^2)\,\omega_z/\omega_0}$.

{\it Large-$N$ field theory.---} For both the closed and open Dicke models, the critical exponents can be analyzed within a unified Keldysh field theory \cite{Torre2013,Sieberer2016,Kamenev2023}, with the leading $1/N$ corrections kept explicitly. The theory is written in terms of the real order-parameter field $x(t)$ associated with the Hermitian operator $\hat x$ defined above. On the Keldysh contour, this field is doubled into $x_+(t)$ and $x_-(t)$ on the forward and backward time branches. After rotating to the Keldysh basis, with $x_{\rm cl}=(x_++x_-)/\sqrt{2}$ and $x_{\rm q}=(x_+-x_-)/\sqrt{2}$, the effective action takes the form
\begin{equation}
\begin{aligned}
	S = -\int dt \,&\Big[x_{\rm q}\left(2\kappa \partial_t + K\partial_t^2 + M r_\kappa\right)x_{\rm cl} \\
	&+ \frac{u}{N}\left(x_{\rm q}x_{\rm cl}^3 + x_{\rm q}^3x_{\rm cl}\right) - {\rm i} D_\kappa x_{\rm q}^2\Big].
\end{aligned}
  \label{eq:action}
\end{equation}
The tuning parameter is the distance from criticality
\begin{equation}
r_\kappa=\frac{(g/g_c)^2-1}{\sqrt{\omega_0/\omega_z+\kappa^2/\omega_0\omega_z}}.
\end{equation}
The noise strength $D_\kappa$ distinguishes the closed and open cases: it vanishes for $\kappa=0$ and is nonzero for $\kappa\neq0$. The remaining coefficients $K$, $M$, and $u$ are nonzero in both cases; their definitions, together with the derivation of the action, are given in the End Matter.

This action provides the starting point for power counting. We assign scaling dimensions with respect to the particle number $N$, rather than a linear size $L$, and adopt the convention $[N]\equiv-1$ and $[t]\equiv-z$. A key feature of Eq.~\eqref{eq:action} is that the near-critical behavior is governed by a \emph{single} tuning parameter $r_\kappa$, rather than by $g$ and $\kappa$ separately, even in the presence of dissipation; see the End Matter for details. The near-critical scaling is therefore controlled by the combination $r_\kappa N^{1/\nu}$, which defines the scaling dimension $[r_\kappa]\equiv1/\nu$.

{\it Closed case: $\kappa=0$.} In the closed model, the two time branches are not coupled by noise, and the classical and quantum fields are treated on equal footing: $[x_{\rm cl}]=[x_{\rm q}]$. At criticality, $r_\kappa=0$, and the coherent kinetic term competes with the leading $1/N$ term. Requiring both terms to be dimensionless gives $\int dt \,x_{\rm q} \partial_t^2 x_{\rm cl} \rightarrow z +2[x_{\rm cl}]=0$, and $\int dt \,\frac{1}{N}\left(x_{\rm q}x_{\rm cl}^3 + x_{\rm q}^3x_{\rm cl}\right) \rightarrow -z+1+4[x_{\rm cl}]=0$.
Hence $[x_{\rm cl}]=[x_{\rm q}]=-1/6$ and $z=1/3$. Away from criticality, the mass term $r_\kappa$ scales in the same way as $\partial_t^2$, so $1/\nu=2z$, giving $\nu=3/2$. This is the all-to-all Ising universality class~\cite{VidalDusuel2006}.

{\it Open case: $\kappa\neq0$.}  In the open model, the noise term couples the two time branches and changes the scaling structure. At long times, the dissipative term $\kappa\partial_t$ is more relevant than the coherent term $\partial_t^2$. The term $x_{\rm q}\partial_t x_{\rm cl}$ then implies $[x_{\rm cl}]=-[x_{\rm q}]$. At criticality, the relevant competition is between the dissipative kinetic term, the noise term, and the leading $1/N$ term $x_{\rm q}x_{\rm cl}^3/N$ (since $x_{\rm cl}$ is more relevant than $x_{\rm q}$). Power counting these terms gives $[x_{\rm cl}]=-[x_{\rm q}]=-1/4$ and $z=1/2$. Finally, the mass term $r_\kappa$ scales in the same way as $\partial_t$, so $1/\nu=z$, giving $\nu=2$. This is the Model-A universality class~\cite{Torre2013}.

The above analysis concerns the static tuning parameter $r_\kappa$. For ramping dynamics, we take $r_\kappa=st$, where $s$ is the ramping speed and define $[s]=1/\mu$; see Fig.~\ref{fig:fig1}(a). Since $[s]=[r_\kappa]-[t]$, we obtain $1/\mu=1/\nu+z$, which is the KZ scaling relation. In the following sections, we recover these large-$N$ critical exponents using the mesoscopic scaling protocol.

{\it Scaling framework at mesoscopic scales.---} 
We now formulate a scaling framework for extracting critical exponents from mesoscopic all-to-all interacting systems. As a reference point, consider a locally interacting system near criticality, with $r$ measuring the distance from the critical point. In the thermodynamic limit, the singular behavior is governed by a diverging correlation length $\xi\sim |r|^{-\nu}$ and correlation time $\tau\sim \xi^z\sim |r|^{-z\nu}$. In a finite system of linear size $L$, this divergence is cut off when $\xi\sim L$, giving the standard scaling variable $rL^{1/\nu}$. If the system is instead driven through the transition at a finite speed, $r=st$, the growth of $\xi$ is cut off by the KZ scale $\xi_{\rm KZ}\sim s^{-\mu}$, where $1/\mu=1/\nu+z$. Thus $\nu$, $z$, and $\mu$ characterize the static, dynamical, and driven scaling properties of the transition, and together provide key fingerprints of its universality class.

For the all-to-all systems considered here, there is no natural notion of a linear length scale. The particle number $N$ therefore replaces $L$ as the finite-size scaling variable. We keep the notation $\nu$, but it should now be understood as the finite-$N$ critical-window exponent: the width of the critical region scales as $N^{-1/\nu}$. A generic observable $Q$ in equilibrium, or in the steady state of an open system, is then expected to obey
$Q(N,r)=N^{y_Q/\nu}\mathcal{F}_Q\!\left(rN^{1/\nu}\right)$,
where $\mathcal{F}_Q$ is the scaling function associated with $Q$~\cite{Cardy1996}. A dimensionless observable has $y_Q=0$ and can be used to extract $\nu$, while a relaxation time has $y_Q/\nu=z$ and can be used to extract $z$.

This scaling form is the natural starting point, but it is not sufficient in the mesoscopic regime. In all-to-all models, the approach to the thermodynamic scaling limit can be slow, so leading irrelevant corrections may strongly affect the accessible sizes and, if ignored, produce apparent exponents that differ substantially from the thermodynamic values. The central idea of our protocol is to first determine the leading irrelevant correction and then use the same correction consistently for other observables [Fig.~\ref{fig:fig1}(c)].

\emph{Stage I: Fix the irrelevant exponent $\omega$.}
We begin with a dimensionless observable $U$, for example a Binder ratio~\cite{LandauBinder2021}. Let $u$ denote the amplitude of the leading irrelevant scaling field, and let $\omega$ be the corresponding correction exponent. The scaling form becomes
\begin{equation}
	\begin{aligned}
		U(N,r) &= \mathcal{F}_U \!\left(rN^{1/\nu},uN^{-\omega}\right)\\
		& = \sum_{j=0} f_{U,j}\!\left(rN^{1/\nu}\right) N^{-j\omega},
	\end{aligned}
	\label{eq:A_expansion}
\end{equation}
where the second line is an expansion in the irrelevant variable, with the coefficients absorbing powers of $u$. At the critical point~\footnote{In practice, the critical point $g_c$ can be determined independently, either numerically from a phenomenological renormalization-group analysis \cite{Fytas2013,BaityJesi2013,Li2025Random} or analytically from large-$N$ theory.}, this reduces to $U(N,0)=\sum_{j=0} f_{U,j}(0)\, N^{-j\omega}$. Thus the finite-$N$ drift of $U(N,0)$ can be fitted to a truncated series in powers of $N^{-\omega}$, allowing $\omega$ to be determined. 

\emph{Stage II: Extract $z$, $\nu$, and $\mu$.}
The remaining exponents are then obtained from independent measurements, all using the same irrelevant correction exponent $\omega$.

$\bullet$\; For $z$: The exponent $z$ is encoded in the relaxation time $\tau$, and hence in the inverse gap $1/\Delta$. The relevant gap is the excitation gap in a Hamiltonian setting or the Liouvillian gap in a dissipative setting~\cite{Minganti2018Liouvillian}. At criticality, it scales as
\begin{equation}
	\Delta(N,0)=N^{-z}\sum_{j=0} f_{\Delta,j}(0) N^{-j\omega},
	\label{eq:fitZ}
\end{equation}
so the size dependence of the critical gap determines $z$.

$\bullet$\; For $\nu$:
Differentiating Eq.~(\ref{eq:A_expansion}) with respect to $r$ and evaluating at criticality gives
\begin{equation}
	\left.\partial_r U(N,r)\right|_{r=0}
	=
	N^{1/\nu}\sum_{j=0} f_{U,j}^{(1)} N^{-j\omega},
	\label{eq:fit_nu}
\end{equation}
where $f_{U,j}^{(1)}$ denotes the first derivative of $f_{U,j}$ at the critical point. Thus the size dependence of the derivative of the dimensionless quantity determines $\nu$~\footnote{In practice, $\partial_r U|_{r=0}$ can be obtained from a symmetric finite difference around $r=0$.}.

$\bullet$\; For $\mu$:
The exponent $\mu$ is extracted from driven dynamics. We consider a linear ramp through the critical point with speed $s$. For a closed system, the initial state is the ground state far from the critical point; for an open system, the corresponding preparation is the steady state. During the ramp, we measure the instantaneous dimensionless quantity $U(N,r,s)$, whose scaling form is~\cite{Gong2010FiniteTimeScaling,DeGrandi2011,Kolodrubetz2012,Liu2014,Huang2014}
\begin{equation}
	U(N,r,s) = \mathcal{F}_U\!\left(rN^{1/\nu},sN^{1/\mu},uN^{-\omega}\right).
	\label{eq:B_expansion}
\end{equation}
We focus on the slow-ramp regime, where the response is controlled by the critical point rather than by fast-quench dynamics. Expanding Eq.~\eqref{eq:B_expansion} in the scaling variable $sN^{1/\mu}$, let $n$ denote the order of the first nonvanishing speed correction. Then
\begin{equation}
	\left. \partial^n_s U(N,r,s) \right|_{r,s=0}\!\!
	= N^{n/\mu} \sum_{j=0} \tilde{f}_{U,j}^{(n)} N^{-j\omega},
	\label{eq:findmu}
\end{equation}
where $\tilde{f}_{U,j}^{(n)}$ denotes the corresponding $n$-th derivative, taken with respect to the speed scaling field, of the $j$-th coefficient in the irrelevant correction expansion. Thus the size dependence of this leading small-speed response determines $\mu$.

This protocol shows that, once $\omega$ is fixed, the exponents $z$, $\nu$, and $\mu$ can be extracted independently. Their agreement with $1/\mu=1/\nu+z$ then provides a nontrivial check that the mesoscopic data analysis recovers the universal KZ scaling, as we verify below for both the closed and open Dicke models.

{\it Numerical results at mesoscopic scales.---}
In the thermodynamic limit, the superradiant transition is characterized by spontaneous $\mathbb{Z}_2$ symmetry breaking and a nonzero order parameter $\mathrm{Tr}(\hat\rho \hat x)\neq0$. At finite $N$, however, the $\mathbb{Z}_2$ symmetry is not truly broken, so the transition must be diagnosed through fluctuations. We use the dimensionless Binder ratio~\cite{King2023DWaveSpinGlass,Wang2025}
\begin{equation}
	U(N)=1-\frac{\mathrm{Tr}(\hat \rho \hat x^4)}{3[\mathrm{Tr}(\hat\rho\hat x^2)]^2}.
	\label{eq:Binder}
\end{equation}
Here $\hat\rho$ can denote the ground-state density matrix $|\mathrm{GS}\rangle\langle\mathrm{GS}|$, the steady state of the Lindblad equation, or a time-dependent state generated by Eq.~\eqref{eq:lindblad}.

\begin{figure}[t]
	\centering
	\safeincludegraphics[width=0.48\textwidth]{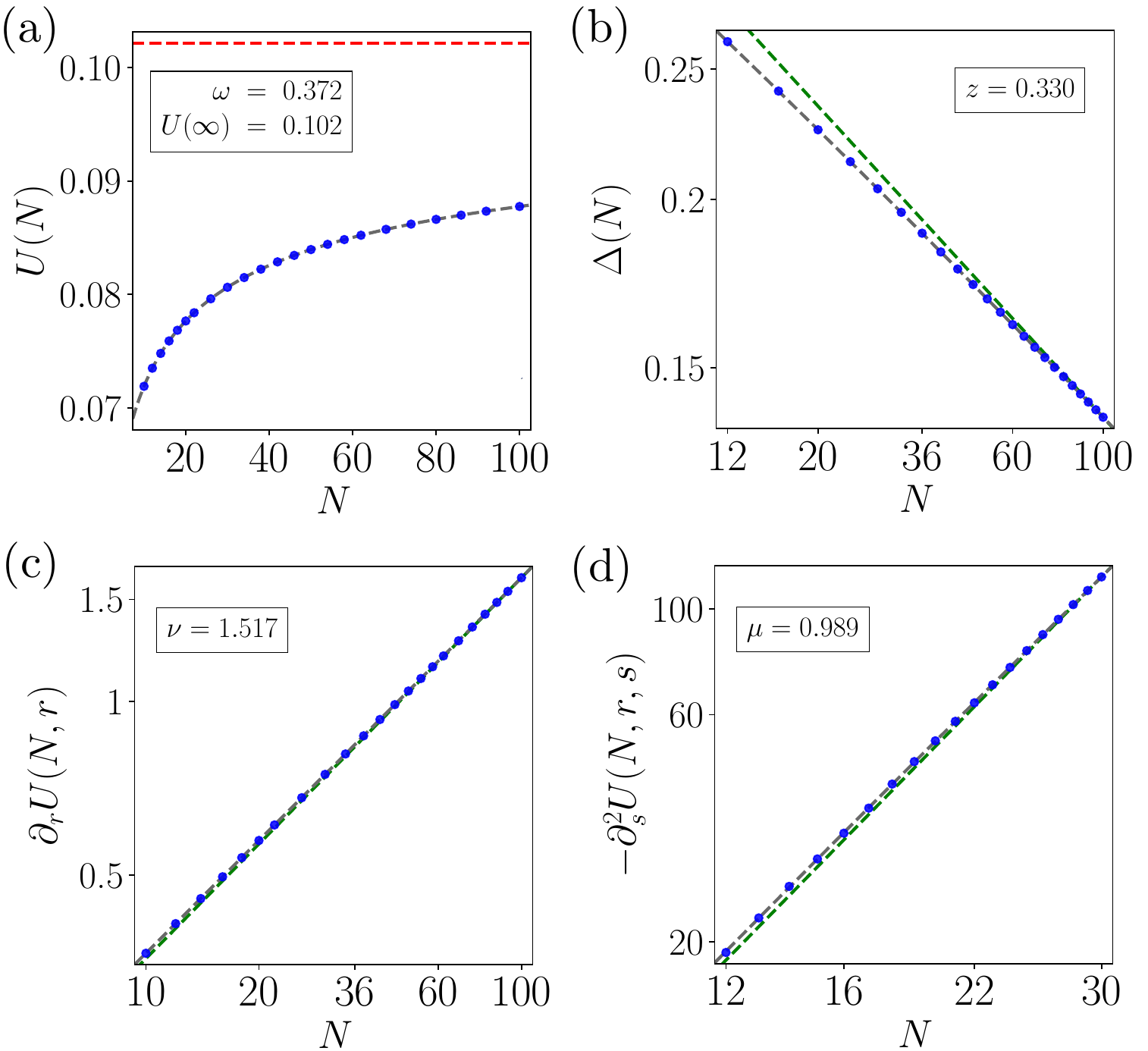}
	\caption{Closed Dicke model. (a) Critical Binder ratio used to fix the leading irrelevant exponent $\omega$. The red dashed line marks extrapolated $U(N\to\infty)$. (b)--(d) Extraction of $z$, $\nu$, and $\mu$ from the critical gap, Binder-ratio derivative, and small-speed response, respectively. Gray dashed curves include corrections to scaling up to $j=2$; green dashed lines show the leading power laws with the fitted exponents. For all figures, we set $\omega_0=\omega_z=1$.}
	\label{fig:fig2}
\end{figure}
		
\begin{figure}[t]
	\centering
	\safeincludegraphics[width=0.48\textwidth]{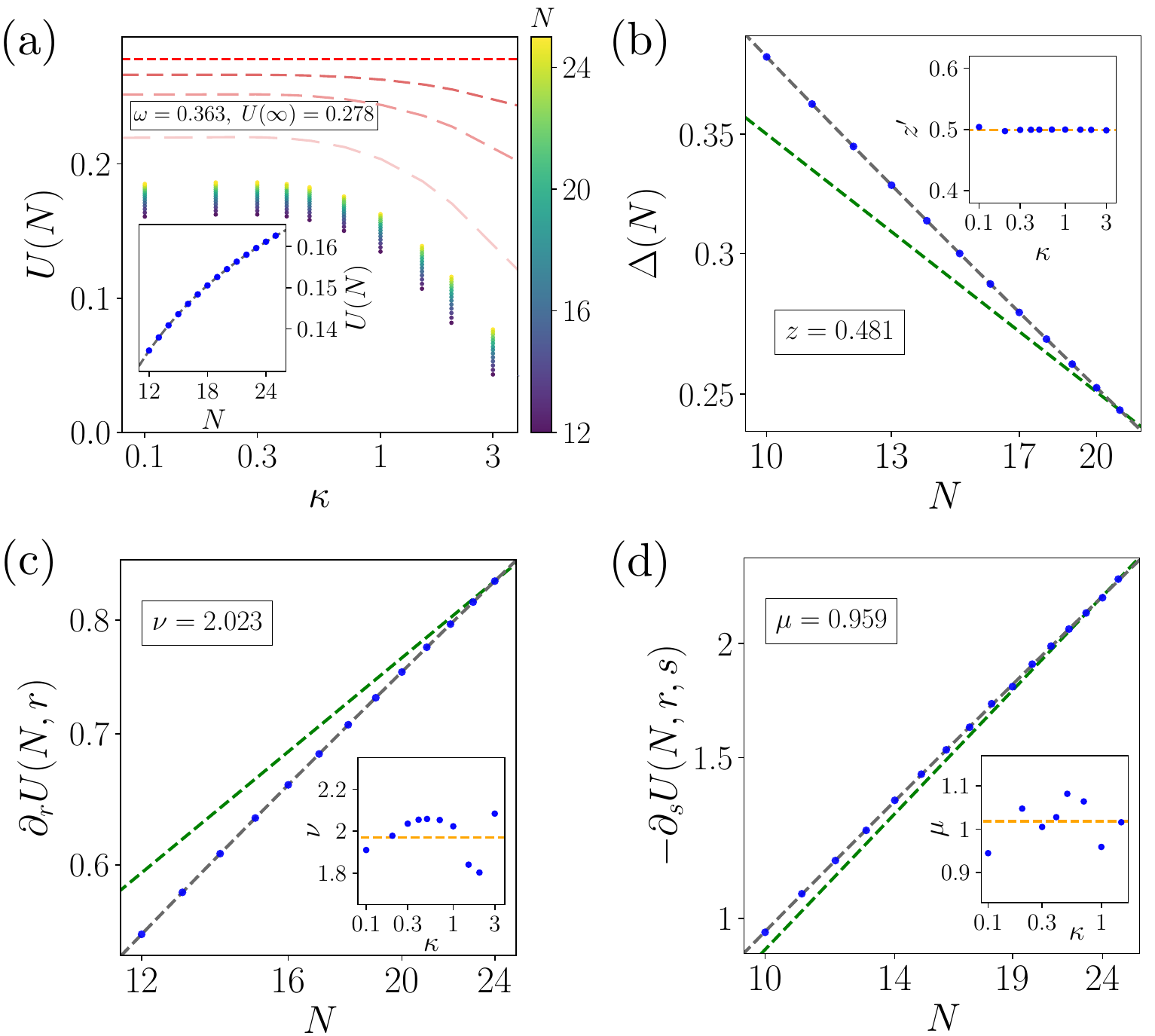}
	\caption{Open Dicke model. (a) Joint critical Binder-ratio fit for different $\kappa\in[0.1,3]$, used to fix $\omega$. Red dashed curves show extrapolated $U(N)$ vs $\kappa$ at fixed $N=10^2,10^3,10^4$, and $N\to\infty$, ordered from sparse to dense. (b)--(d) Extraction of $z$, $\nu$, and $\mu$ from the Liouvillian gap, Binder-ratio derivative, and linear small-speed response, respectively. Insets in (b)--(d) show the corresponding $\kappa$ dependence.}
	\label{fig:fig3}
\end{figure}

{\it Closed case:}
We first consider the ground state. Figure~\ref{fig:fig2}(a) shows the finite-$N$ dependence of the Binder ratio at the critical point $r=0$. Fitting the drift of $U(N)$ with Eq.~\eqref{eq:A_expansion}, truncated to $j=2$ in the irrelevant correction, gives $\omega=0.372$ using sizes up to $N=100$. The extrapolated infinite-$N$ value, $U(N\rightarrow\infty)\approx0.102$, is shown by the red dashed line in Fig.~\ref{fig:fig2}(a). Its visible separation from the $N=100$ data highlights the strong finite-size corrections in this mesoscopic regime. With this $\omega$ fixed, the critical gap gives $z=0.33$ [Fig.~\ref{fig:fig2}(b)], while the critical derivative of the Binder ratio gives $\nu=1.52$ [Fig.~\ref{fig:fig2}(c)]. These values recover the thermodynamic exponents to two significant digits.

We next study ramping dynamics by driving $r$ from deep in the normal phase through the critical point with speed $s$. For the closed system, time-reversal symmetry forbids a linear-in-$s$ correction to $U(s)$. The leading speed dependence is therefore quadratic. We compute $\partial^2_s U$ and fit its size dependence. This gives $\mu=0.99$, close to the large-$N$ prediction $\mu=1$.

{\it Open case:} 
For the open system, we first analyze the steady state. Although the loss rate $\kappa$ can vary, all $\kappa\neq0$ cases are expected to belong to the same universality class. We therefore impose that both $\omega$ and the critical value of $U(N\rightarrow\infty)$ are independent of $\kappa$. A joint fit over $\kappa\in[0.1,3]$ gives $\omega=0.363$, as shown in Fig.~\ref{fig:fig3}(a).

The dynamical exponent $z$ is obtained from the Liouvillian gap. For $\kappa=1$, exact diagonalization gives $z=0.48$. For generic $\kappa$, however, the small sizes accessible to exact diagonalization, $N\leq24$, show crossings and rearrangements among low-lying Liouvillian modes, which obscure the asymptotic scaling. To avoid these crossover effects, we perform a numerical large-$N$ expansion, described in the End Matter. For $N\in[10^5,10^7]$, the extracted exponent approaches $z=1/2$ for different $\kappa$, as shown in the inset of Fig.~\ref{fig:fig3}(b). The exponent $\nu$ is extracted from the critical derivative of the Binder ratio. At $\kappa=1$, we find $\nu=2.02$ [Fig.~\ref{fig:fig3}(c)], while fits at different $\kappa$ fluctuate around the expected value, giving $\nu=1.97(9)$ in the inset.

For ramping dynamics, because dissipation breaks the time-reversal structure, the Binder ratio can acquire a linear correction in the ramp speed. We therefore compute $\partial_s U$ and fit its size dependence. At $\kappa=1$, we find $\mu=0.96$; across different $\kappa$, we obtain $\mu=1.02(5)$.

\begin{figure}[t]
	\centering
	\safeincludegraphics[width=0.48\textwidth]{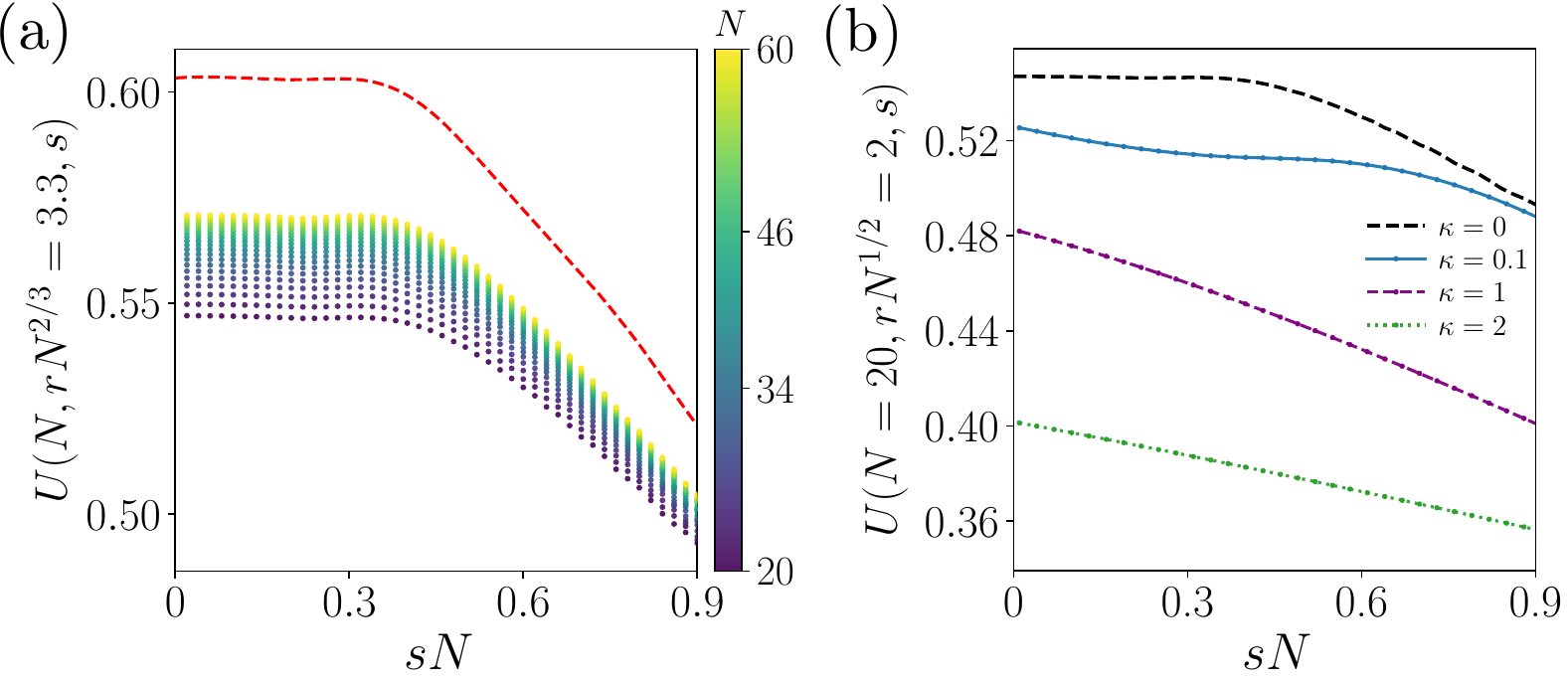}
	\caption{(a) Ramped Binder ratio in the closed Dicke model for different system sizes; red dashed lines indicate the $N\to\infty$ extrapolation. (b) Comparison of the ramped Binder ratio between the closed and open Dicke models for different $\kappa$.
}
	\label{fig:fig4}
\end{figure}

{\it Time scales and crossover.---}
We now discuss how the finite size $N$, the ramp speed $s$, and the dissipation $\kappa$ compete in the ramping dynamics [Fig.~\ref{fig:fig1}(b)]. In the closed model, the relaxation time diverges as $\tau\sim |r|^{-1/2}$ [cf. $z\nu=1/2$]. For a linear ramp $r=st$, the KZ time $t_{\rm KZ}$ is defined by $\tau(r_{\rm KZ})\sim |r_{\rm KZ}|/s$ \cite{Zurek1985}, which gives $t_{\rm KZ}\sim s^{-1/3}$. On the other hand, the critical finite-size gap scales as $\Delta_N\sim N^{-1/3}$, yielding the finite-size time scale $t_N\sim N^{1/3}$. Comparing $t_{\rm KZ}$ with $t_N$ gives the scaling variable $sN$ [cf. $\mu=1$]. Thus $sN\ll1$ is the finite-size-dominated perturbative regime, while $sN\gtrsim1$ marks the onset of the conventional KZ regime. Figure~\ref{fig:fig4}(a) mainly probes the former: the response is nearly flat at small $sN$, consistent with the absence of a linear-in-$s$ correction.

We next consider how $\kappa$ drives the system away from the closed fixed point. In the large-$N$ long-time limit, any finite $\kappa$ is a relevant perturbation: the coherent closed fixed point is unstable, and the dynamics flows to the dissipative fixed point, with $z$ changing from $1/3$ to $1/2$. Near the closed fixed point, the competition between the coherent kinetic term and photon loss yields $\partial_t^2\sim \kappa\,\partial_t$, which defines a dissipative timescale $t_\kappa\sim \kappa^{-1}$. In a finite system, the closed critical dynamics is cut off at $t_N\sim N^{1/3}$. Hence, the crossover scaling variable is $\kappa N^{1/3}$: small systems remain closed-like when $\kappa N^{1/3}\ll1$ and enter the dissipative regime otherwise. This is consistent with scaling theory, where the dissipation rate enters as an additional scaling variable through $Q\sim\mathcal{F}_Q(r_0N^{2/3},\kappa N^{1/3})$~\cite{Yin2014,RossiniVicari2020}. However, the scaling protocol should not be limited to this weak-dissipation crossover. Since the limit $\kappa\to0$ is singular, accessing the dissipative fixed point itself requires a \emph{separate} scaling analysis. This is precisely what our protocol achieves: it extracts the distinct dissipative fixed-point scaling form $\mathcal{F}_Q(r_\kappa N^{1/2})$ [cf.~Fig.~\ref{fig:fig3}(c)].

For ramping at nonzero $\kappa$, the KZ time $t_{\rm KZ}\sim s^{-1/3}$ enters. The same comparison gives $t_{\rm KZ}\sim t_\kappa$, or $s\sim \kappa^3$. Thus fast ramps with $s\gg\kappa^{3}$ do not give the system enough time to relax and can appear closed-like, while slow ramps cross over to open dissipative scaling. This behavior is visible in Fig.~\ref{fig:fig4}(b). For large $\kappa$, such as $\kappa=1,2$, the response is already open-like and shows a linear small-speed decay. For small $\kappa$, such as $\kappa=0.1$, the slowest ramps exhibit the open linear response, but increasing $s$ eventually bends the curve toward the flatter, closed-like speed dependence seen in Fig.~\ref{fig:fig4}(a). These intertwined crossovers explain why the raw ramp data can look complicated, and why a unified finite-size scaling treatment that includes the irrelevant corrections is needed to extract the correct exponents.

{\it Outlook.---}
This framework is not tied to cavity-QED platforms. It applies more broadly to long-range and all-to-all quantum simulators, from trapped ions with tunable power-law interactions~\cite{Li2023} to Sachdev--Ye--Kitaev-type models related to non-Fermi-liquid criticality \cite{Wang2020}. In many such systems, a controlled large-$N$ theory may not be available, so extracting the leading irrelevant structure directly from data becomes essential. Our results therefore provide a practical route for probing universality when the thermodynamic limit is clear in principle but experimentally out of reach.

{\it Acknowledgments.---}
We are grateful to Hui Zhai for stimulating ideas. We also thank Chengshu Li for reading the manuscript and providing valuable feedback. H.L. is supported by Quantum Science and Technology-National Science and Technology Major Project (Grant No. 2025ZD0300400), China National Postdoctoral Program for Innovative Talents (Grant No. BX2026034) , and National Natural Science Foundation of China (Grant No. 12547168). H.W. is supported by China Postdoctoral Science Foundation under Grant No.~2024M751609 and Postdoctoral Fellowship Program of CPSF under Grant No.~GZC20231364.

\bibliography{references_KZ}

\onecolumngrid
\newpage
\twocolumngrid

\section*{End Matter}
\setcounter{equation}{0}
\renewcommand{\theequation}{A\arabic{equation}}
\renewcommand{\theHequation}{A\arabic{equation}}

{\it Appendix A: Keldysh action.---}
Here we derive Eq.~\eqref{eq:action} of the main text. We represent the collective spin degrees of freedom by a Holstein--Primakoff boson $\hat b$,
\begin{equation}
\begin{aligned}
\hat J_z &= \hat b^\dagger\hat b-N/2,\\
\hat J_x &= \frac{1}{2} \left[\hat b^\dagger\sqrt{N-\hat b^\dagger \hat b} + \sqrt{N-\hat b^\dagger \hat b}\,\hat b\right].
\end{aligned}
\end{equation}
Substituting this representation into the Dicke Hamiltonian and expanding to leading order in $1/N$, we obtain $\hat H=\hat H_2+\hat H_{1/N}$, with
\begin{equation}
\begin{aligned}
&\hat H_2 = \omega_0\hat a^\dagger\hat a +\omega_z\hat b^\dagger\hat b +\frac{g}{2}(\hat a+\hat a^\dagger) (\hat b+\hat b^\dagger),\\
&\hat H_{1/N} = -\frac{g}{4N}(\hat a+\hat a^\dagger)(\hat b^\dagger\hat b\hat b+\hat b^\dagger\hat b^\dagger\hat b).
\end{aligned}
\label{eq:em_hp_h}
\end{equation}

We now formulate the theory on the Keldysh contour. Including the Lindblad jump term from Eq.~\eqref{eq:lindblad}, the action reads
\begin{equation}
\begin{aligned}
S=\!\int \! dt\,&\Big[\mathrm{i}(\bar a_+\partial_t a_+ -\bar a_-\partial_t a_-)+\mathrm{i}(\bar b_+\partial_t b_+ -\bar b_-\partial_t b_-)\\
&-(H_+-H_-)-\mathrm{i}\kappa(2a_+\bar a_- -\bar a_+a_+ -\bar a_-a_-)\Big].
\end{aligned}
\label{eq:em_lindblad_contour_action}
\end{equation}
To proceed, we introduce real coordinate and momentum variables for the cavity mode, $x=(a+\bar a)/\sqrt{2\omega_0}$ and $p=\mathrm{i}\sqrt{\omega_0/2}\,(\bar a-a)$. We similarly introduce the atomic coordinate $y=(b+\bar b)/\sqrt{2\omega_z}$ and its conjugate momentum $q=\mathrm{i}\sqrt{\omega_z/2}\,(\bar b-b)$.

We first treat the quadratic part $S_2$. Performing the Gaussian integration over the momenta $p$ and $q$ gives
\begin{equation}
\begin{aligned}
S_2 = &-\int dt\,\left[x_{\rm q}(\partial_t^2+2\kappa\partial_t+\omega_0^2+\kappa^2)x_{\rm cl}
-\mathrm{i}D_\kappa x_{\rm q}^2\right]\\
& -(S_y^+ - S_y^-) .
\end{aligned}
\end{equation}
In the first line, the Keldysh rotation for $x$ follows the convention used in the main text. We have omitted the term $\frac{\mathrm{i}\kappa}{\omega_0}(\partial_t x_{\rm q})^2$, which is irrelevant at low frequencies. The second line contains the remaining action for the massive atomic coordinate $y$, with
\begin{equation}
S_y = \int dt \, \left[\frac{1}{2}y(\partial_t^2+\omega_z^2)y + g\sqrt{\omega_0\omega_z} \,xy\right].
\label{eq:em_Sy}
\end{equation}

We now turn to the $1/N$ contribution, which supplies the leading interaction vertex. In terms of $x$, $y$, and $q$, the interaction reads
\begin{equation}
H_{1/N} = -\frac{g\sqrt{\omega_0}\,\omega_z^{3/2}}{4N}xy^3 -\frac{g}{4N} \sqrt{\frac{\omega_0}{\omega_z}} xyq^2 .
\end{equation}
The $q^2$ term gives only irrelevant contributions to the critical theory, so the $xy^3/N$ term provides the leading relevant vertex.

We are now ready to integrate out the massive field $y$. Without the $1/N$ correction, one can eliminate $y$ by a Gaussian integration. With the $1/N$ vertex present, it is useful to first shift $y\rightarrow y-\mathcal{G}^{-1}x$, so that the shifted $y$ field has zero mean. The quadratic action \eqref{eq:em_Sy} becomes
\begin{equation}
S_y \rightarrow \int dt \, \left[\frac{1}{2}y(\partial_t^2+\omega_z^2)y\right]-\int dt \,\left[\frac{1}{2}g\sqrt{\omega_0\omega_z} \,x\mathcal{G}^{-1}x\right],
\label{eq:em_Sy_shift}
\end{equation}
where, in the low-frequency limit,
\begin{equation}
\mathcal{G}^{-1} = g\sqrt{\omega_0\omega_z}(\partial_t^2+\omega_z^2)^{-1} \approx g\sqrt{\frac{\omega_0}{\omega_z^3}}(1-\omega_z^{-2}\partial_t^2).
\end{equation}
This shift removes the bilinear $xy$ coupling. The second term in \eqref{eq:em_Sy_shift} renormalizes the coherent kinetic and mass terms of the cavity field.

We then treat the $xy^3/N$ vertex perturbatively using the shifted zero-mean Gaussian field $y$,
\begin{equation}
S_{1/N} = -\int dt\,\frac{g\sqrt{\omega_0}\,\omega_z^{3/2}}{4N}\left\langle x(y-\mathcal{G}^{-1}x)^3 \right\rangle_y ,
\end{equation}
where $\langle\cdot\cdot\cdot\rangle_y$ denotes averaging over the zero-mean Gaussian action. The terms $x\langle y^3\rangle/N$ and $x^3\langle y\rangle/N$ vanish, while $x^2\langle y^2\rangle/N$ gives an $\mathcal{O}(1/N)$ mass renormalization and shifts the critical point. The only relevant nonlinear term is therefore the quartic contribution $x^4/N$. Expressed on the forward--backward contour as $x_+^4-x_-^4$ and rotated to the Keldysh basis, it gives
\begin{equation}
S_{1/N} \approx -\int dt\,\frac{u}{N} \left(x_{\rm q}x_{\rm cl}^3+x_{\rm q}^3x_{\rm cl}\right), \quad u=\frac{g^4\omega_0^2}{2\omega_z^3}.
\end{equation}

Collecting the quadratic renormalizations, the effective quadratic action takes
the form
\begin{equation}
S_2=-\int dt\,\left[x_{\rm q} \left(2\kappa\partial_t+K\partial_t^2+ \widetilde{M} \right)x_{\rm cl}-\mathrm{i}D_\kappa x_{\rm q}^2\right],
\end{equation}
with renormalized coefficients
\begin{equation}
K=1+\frac{g^2\omega_0}{\omega_z^3},\quad
\widetilde{M} = -(\omega_0^2+\kappa^2) \left[(g/g_c)^2-1\right].
\end{equation}
Near the critical point, $\widetilde{M}$ is linear in $g-g_c(\kappa)$ and controls the transition. The coefficients $K$ and $u$ are nonsingular at criticality and can be evaluated at $g=g_c(\kappa)$.

\setcounter{equation}{0}
\renewcommand{\theequation}{B\arabic{equation}}
\renewcommand{\theHequation}{B\arabic{equation}}

{\it Appendix B: Universal theory for $\kappa\neq0$.---}
We now show that, for any finite loss rate $\kappa$, the low-energy dissipative theory can be written in terms of a \emph{single} scaled distance from criticality. The key observation is that the explicit $\kappa$ dependence of the quadratic and nonlinear coefficients can be absorbed entirely into a rescaling of the fields and time.

In the dissipative regime, the damping term $\kappa\partial_t$ dominates over $\partial_t^2$ at low frequencies. We therefore drop $\partial_t^2$ and perform the rescaling
\begin{equation}
\begin{aligned}
\begin{cases}
t=\frac{8\kappa N^{1/2}\omega_z^2}{g_c^3(\kappa)\omega_0}\,\tilde{t},\\
x_{\rm cl}(t)=N^{1/4}
\left(\frac{2\omega_z}{g_c(\kappa)\omega_0}\right)^{1/2}\phi_{\rm cl}(\tilde{t}),\\
x_{\rm q}(t)=\frac{1}{2\kappa N^{1/4}}
\left(\frac{g_c(\kappa)\omega_0}{2\omega_z}\right)^{1/2}\phi_{\rm q}(\tilde{t}).
\end{cases}
\end{aligned}
\end{equation}
With this choice, the effective action becomes
\begin{equation}
S = -\!\int \! d\tilde{t} \left[\phi_{\rm q}\!\left(\partial_{\tilde{t}}-4r_\kappa N^{1/2}\right)\!\phi_{\rm cl} + 4\phi_{\rm q}\phi_{\rm cl}^3 -\mathrm{i}\phi_{\rm q}^2 \right],
\end{equation}
where the scaled distance from criticality is
\begin{equation}
r_\kappa=\frac{(g/g_c)^2-1}{\sqrt{\omega_0/\omega_z+\kappa^2/\omega_0\omega_z}}.
\end{equation}
Thus the dissipative low-energy theory depends on the microscopic parameters only through the scaling variable $r_\kappa N^{1/2}$. This gives the open-system exponent $\nu=2$. Hence, the corresponding dissipative fixed-point scaling takes the form $Q\sim \mathcal{F}_Q(r_\kappa N^{1/2})$. This should be distinguished from the weak-dissipation crossover scaling near the closed fixed point, $Q\sim \mathcal{F}_Q(r_0N^{2/3},\kappa N^{1/3})$, where $r_0=(g/g_c)^2-1$. 

This rescaling also clarifies the form of the mass term in Eq.~\eqref{eq:action}. The bare mass of Appendix~A factorizes as $\widetilde{M}=M r_\kappa$, isolating the scaling variable $r_\kappa$, with
\begin{equation}
M=-g_c^3(\kappa)\frac{\omega_0}{\omega_z^2}.
\end{equation}
	
\setcounter{equation}{0}
\renewcommand{\theequation}{C\arabic{equation}}
\renewcommand{\theHequation}{C\arabic{equation}}
{\it Appendix C: Hartree--Fock--Bogoliubov analysis of the open-system soft mode.---}
At the small sizes accessible to exact Liouvillian diagonalization, the lowest nonzero eigenmode can switch between different branches as $N$ and $\kappa$ are varied. 
This obscures the asymptotic scaling of the critical soft mode.

We treat the $1/N$ interaction in Eq.~\eqref{eq:em_hp_h} at the HFB level. 
With $n_b=\langle\hat b^\dagger\hat b\rangle$ and $m_b=\langle\hat b\hat b\rangle$,
\begin{equation}
    \hat b^\dagger\hat b^\dagger\hat b
    +\hat b^\dagger\hat b\hat b
    \simeq
    (2n_b+\mathrm{Re}\,m_b)
    (\hat b+\hat b^\dagger).
    \label{eq:em_hfb_decouple}
\end{equation}
The effective Hamiltonian therefore becomes
\begin{equation}
    \hat H_{\rm HFB}
    =
    \omega_0\hat a^\dagger\hat a
    +\omega_z\hat b^\dagger\hat b
    +\frac{g_{\rm eff}}{2}
    (\hat a+\hat a^\dagger)(\hat b+\hat b^\dagger),
    \label{eq:em_hfb_hamiltonian}
\end{equation}
where $g_{\rm eff}=g[
1-(2n_b+\mathrm{Re}\,m_b)/2N].$
Together with the cavity loss in Eq.~\eqref{eq:lindblad}, this gives
\begin{equation}
    \dot{\hat v}=A\hat v+\hat \xi ,
    \qquad
    \hat v=(\hat a,\hat a^\dagger,\hat b,\hat b^\dagger)^T.
    \label{eq:em_hfb_drift}
\end{equation}
The drift matrix is
\begin{equation}
\begin{aligned}
    A=
    \begin{pmatrix}
        -\kappa-\mathrm{i}\omega_0 & 0 & -\mathrm{i}g_{\rm eff}/2 & -\mathrm{i}g_{\rm eff}/2\\
        0 & -\kappa+\mathrm{i}\omega_0 & \mathrm{i}g_{\rm eff}/2 & \mathrm{i}g_{\rm eff}/2\\
        -\mathrm{i}g_{\rm eff}/2 & -\mathrm{i}g_{\rm eff}/2 & -\mathrm{i}\omega_z & 0\\
        \mathrm{i}g_{\rm eff}/2 & \mathrm{i}g_{\rm eff}/2 & 0 & \mathrm{i}\omega_z
    \end{pmatrix}.
\end{aligned}
    \label{eq:em_hfb_A}
\end{equation}
The input noise satisfies,
\begin{equation}
    \langle \hat \xi(t)\hat \xi^\dagger(t')\rangle
    =D\delta(t-t'),
    \quad
    D=\mathrm{diag}(2\kappa,0,0,0).
    \label{eq:em_hfb_D}
\end{equation}
The covariance matrix $C_{ij}=\langle\hat v_i\hat v_j^\dagger\rangle$ satisfies
\begin{equation}
    AC+CA^\dagger+D=0 .
    \label{eq:em_hfb_lyapunov}
\end{equation}
The moments $n_b=\langle\hat b^\dagger\hat b\rangle$ and $m_b=\langle\hat b\hat b\rangle$ extracted from $C$ are inserted back into $g_{\rm eff}$, and this loop is iterated to self-consistency.
\begin{figure}[b]
    \centering
    \safeincludegraphics[width=0.42\textwidth]{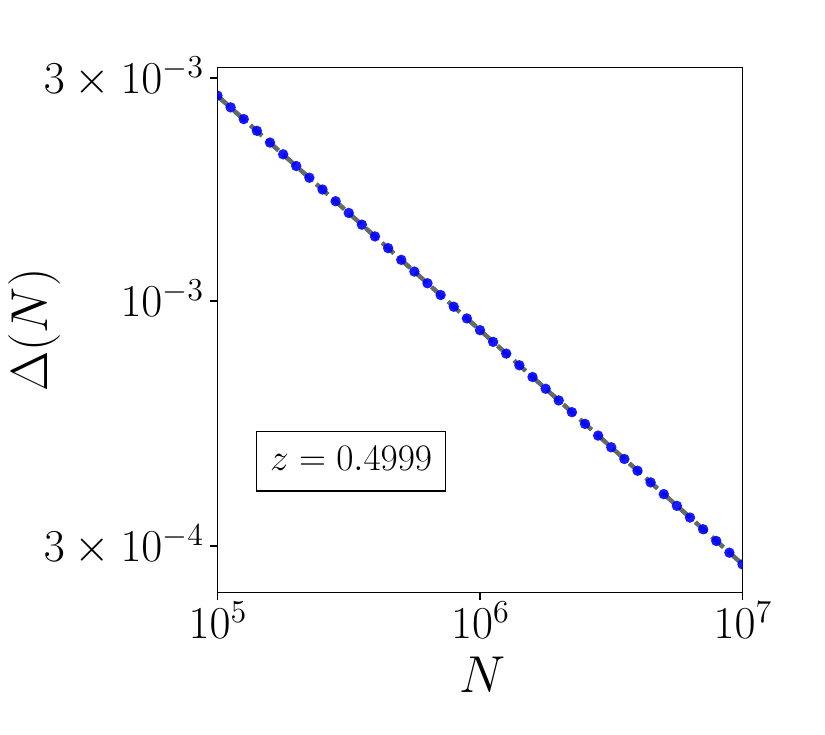}
    \caption{Soft-mode spectral gap of the open Dicke model at the critical point, obtained within the Hartree--Fock--Bogoliubov approximation for $\kappa=1$. The calculation is performed over the range $N=10^5$--$10^7$, where the asymptotic $N^{-1/2}$ decay becomes clearly visible.}
    \label{fig:em_hfb_gap}
\end{figure}

After convergence, the homogeneous equation has modes $\delta\hat v(t)\propto e^{\lambda t}$, where $\lambda$ are eigenvalues of $A$. 
The soft-mode decay rate is therefore
\begin{equation}
    \Delta_{\rm soft}
    =
    -\max_{\mathrm{Re}\,\lambda<0}
    \mathrm{Re}\,\lambda .
    \label{eq:em_hfb_gap}
\end{equation}
This quantity equals the Liouvillian gap for a quadratic open bosonic model. 
Here it is used only as an estimator of the asymptotic critical branch. 
As shown in Fig.~\ref{fig:em_hfb_gap}, $\Delta_{\rm soft}\sim N^{-1/2}$ for $N=10^5$--$10^7$, confirming the dissipative exponent $z=1/2$ used in the main text.

\end{document}